\documentclass[aps,twocolumn,superscriptaddress,hyperref]{revtex4}       
\usepackage{graphicx,psfrag,amsmath,times,hhline,graphicx,color}
\begin{document}       
\title{Electronic circuit implementation of chaos synchronization } 
\author{Ravi Ranjan}
\author{Shivshankar Mishra}
\author{Suneel Madhekar }
\affiliation{PGAD, Defence Research and Development Organisation, Kanchanbagh, Hyderabad - 500 058, India.}

\begin{abstract}
In this paper, an electronic circuit implementation of a robustly chaotic two-dimensional map is presented.  Two such electronic circuits are realized.  One of the circuits is configured as the driver and the other circuit is configured as the driven system.  Synchronization of chaos between the driver and the driven system is demonstrated.
\end{abstract}
\maketitle
\section{Introduction}         
Chaotic systems are sensitive to initial conditions and diverge exponentially. Even if there is an infinitesimally small difference in initial conditions, two identical chaotic systems spiral out along completely different paths. Synchronization of chaotic dynamics is a phenomenon wherein two or more chaotic systems, in the presence of a drive signal, adjust some property of their dynamics to bear a mathematical relationship.  According to Pecora and Carroll, chaotic flows or iterated maps can be synchronized, in such a way that their states are almost identical, if they are coupled with a drive signal \cite{Pecora_1990}.

Robustness of chaos implies that in some neighborhood of the parameter space, the chaotic dynamics is free of any windows of periodicity. The absence of any windows of periodicity ensures that small changes in the parameter do not destroy chaos. Such structural stability of chaos is valuable, especially in engineering applications \cite{robust}. 

In \cite{Pecora_1990} synchronization was achieved between electronic circuit realization of two flows.  The flows were modified versions of an RC op-amp chaos generator. Few electronic circuit realizations of iterative chaotic maps have been reported in \cite{SuneelM_2006} and \cite{Campos_2009}. In this paper, an analog electronic circuit realization of a robustly chaotic iterative two-dimensional map is presented. Then, synchronization between two such electronic circuits is demonstrated. 
\section{The Map and the coupling scheme}
\begin{figure}
\scalebox{.5}{\input{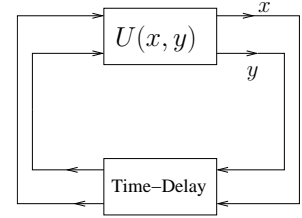}}
\caption{Synchronization scheme}
\label{map}
\end{figure}

\begin{figure}
\scalebox{.5}{\input{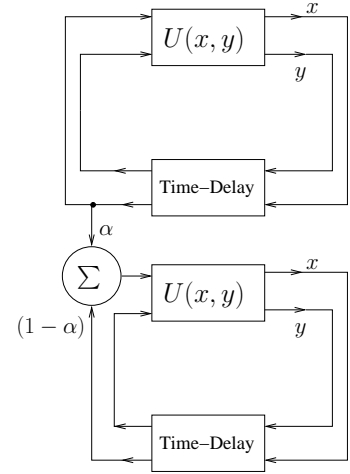}}
\caption{Coupling scheme}
\label{sync}
\end{figure}

The implemented map is a two-dimensional chaotic system that reduces to the H\'{e}non and Lozi systems at its extremes \cite{Elhadj_2007}. The map is defined as
\begin{equation}
(x_{k+1},y_{k+1})^T=U\left(x_k,y_k\right) ,
\end{equation}
where,
\begin{equation}
\label{eqnElhadj}
U(x,y)={{1-1.4f_{\alpha _b}(x)+y}\choose {0.3x}} ,
\end{equation}
and $\left(x_k, y_k\right)$ is the state of the system.
Here, $\alpha _b$ is the bifurcation parameter and
\begin{equation}
f_{\alpha _b}\left(x\right)=\alpha _b\left |  x \right | + \left(1-\alpha _b\right)x^2 .
\end{equation}
For $\alpha _b=1$ and $\alpha _b=0$, (\ref{eqnElhadj}) reduces to the Lozi map and the H\'{e}non map, respectively. The map is robustly chaotic for $0 \leq \alpha \leq 1$. The implementation scheme is shown in figure \ref{map}.
In this paper, the drive signal $x_k$ of the driver system is uni-directionally coupled to the driven system. The driver system can be written as,
\begin{equation}
\left(x_{k+1}^r,y_{k+1}^r\right)^T=U\left(x_k^r,y_k^r\right) , 
\end{equation}
where, $\left( x_k^r, y_k^r \right)$ denotes the state of the driver system at time $k$. The driven system can be written as,
\begin{equation}
\left(x_{k+1}^n,y_{k+1}^n\right)^T=U\left(\alpha _c x_k^r+\left(1-\alpha _c\right)x_k^n,y_k^n\right) ,
\end{equation}
where, $\left( x_k^n, y_k^n \right)$ denotes the state of driven system at time $k$. $0 \leq \alpha _c \leq 1$ is the coupling factor between the two systems. The coupling scheme is illustrated in figure \ref{sync}.

\section{Circuit realization}
A block diagram representation of the electronic circuit realization is shown in figure \ref{block}. Here, the bifurcation factor, $\alpha_{b}$ is chosen to be 0.4. The upper block of figure \ref{block} implements $U(x,y)$ and computes the next state of the system, given the present state at its input. The lower block of figure \ref{block} is a pair of sample-and-hold blocks, that remember the state of system for the duration of one clock cycle.
\begin{figure}
\scalebox{.75}{\input{block.pstex_t}}
\caption{Circuit implementation block diagram}
\label{block}
\end{figure}
 
\subsection{The circuit}
The complete circuit used to realize the blocks of figure \ref{block}, is shown in figure \ref{ckt}.  In the reported circuit realization, for computing $\left| x \right|$, for multiplying with a constant, and for a weighted sum, operational amplifiers are used.  The operational amplifiers used were the integrated circuit $LM324$. The function $x^2$ is implemented by an analog multiplier, $MPY634$ \cite{MPY634}. Each sample-and-hold circuit is implemented using one $LF398$ integrated circuit \cite{LF398}.  The $LF398$ is a level-triggered circuit.  Therefore, to obtain the necessary discrete-time nature of the map, each time-delay block is implemented in the form of two sample-and-hold circuits.  The two circuits are driven by clock-waveforms that are $180^{o}$ out-of-phase.  

\subsection{Synchronization}

The synchronization scheme used is the same as that shown in figure \ref{sync}. Here the state variable $x_k$ of driver circuit, is fed to the driven circuit, through a coupling factor.
\begin{figure}
\scalebox{.5}{\input{ckt.pstex_t}}
\caption{Circuit Diagram of Map Implementation}
\label{ckt}
\end{figure}
\begin{figure}
\scalebox{.5}{\input{sync_ckt.pstex_t}}
\caption{Circuit Diagram of Map Synchronization}
\label{sync_ckt}
\end{figure}

The coupling circuit is just a weighted sum circuit, followed by a voltage follower stage, as shown in figure \ref{sync_ckt}. The output of coupling circuit is $\alpha_c x_k^r + \left( 1- \alpha_c \right) x_k^n$. For $\alpha_c = 0.5$, $R_f = 5 K\Omega, R_n = 10 K \Omega$ and $R_r = 10 K \Omega$. To obtain the value of of $5K\Omega$ two $10K\Omega$ resistors were connected in parallel. For $\alpha_c=0.4$; $R_f = 3.3 K \Omega, R_n = 8 K \Omega$ and $R_r = 5.5 K \Omega$. To obtain the value of $8K\Omega$, a $4.7K\Omega$ resistor and a $3.3K\Omega$ resistor were used in series. To obtain the value of $5.5K\Omega$, a $2.2K\Omega$ resistor and a $3.3K\Omega$ resistor were used in series.

\section{Results}
\begin{figure}
\includegraphics[scale=.25]{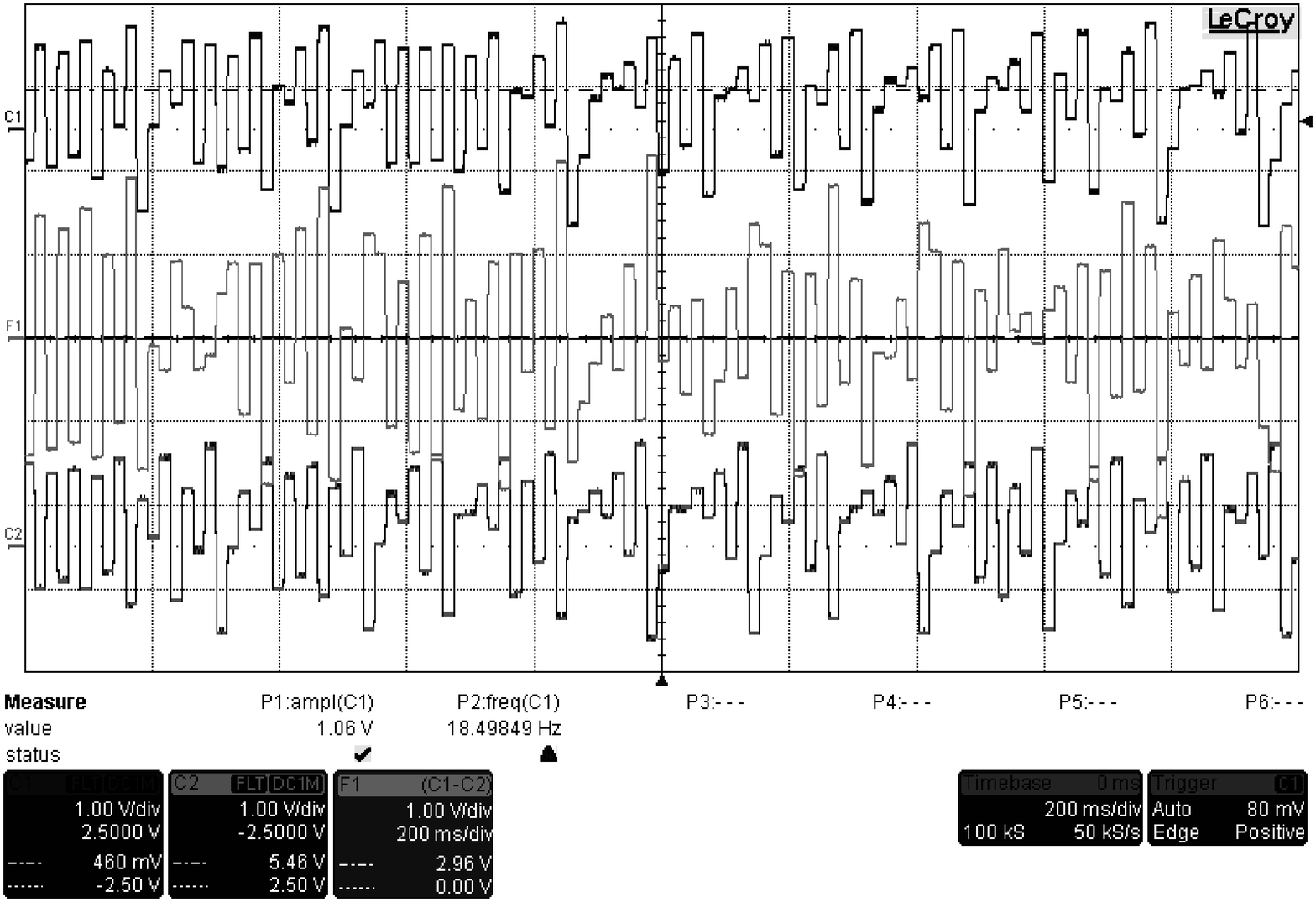}
\caption{No Coupling}
\label{UncoupledTrace}
\end{figure}

\begin{figure}
\includegraphics[scale=.25]{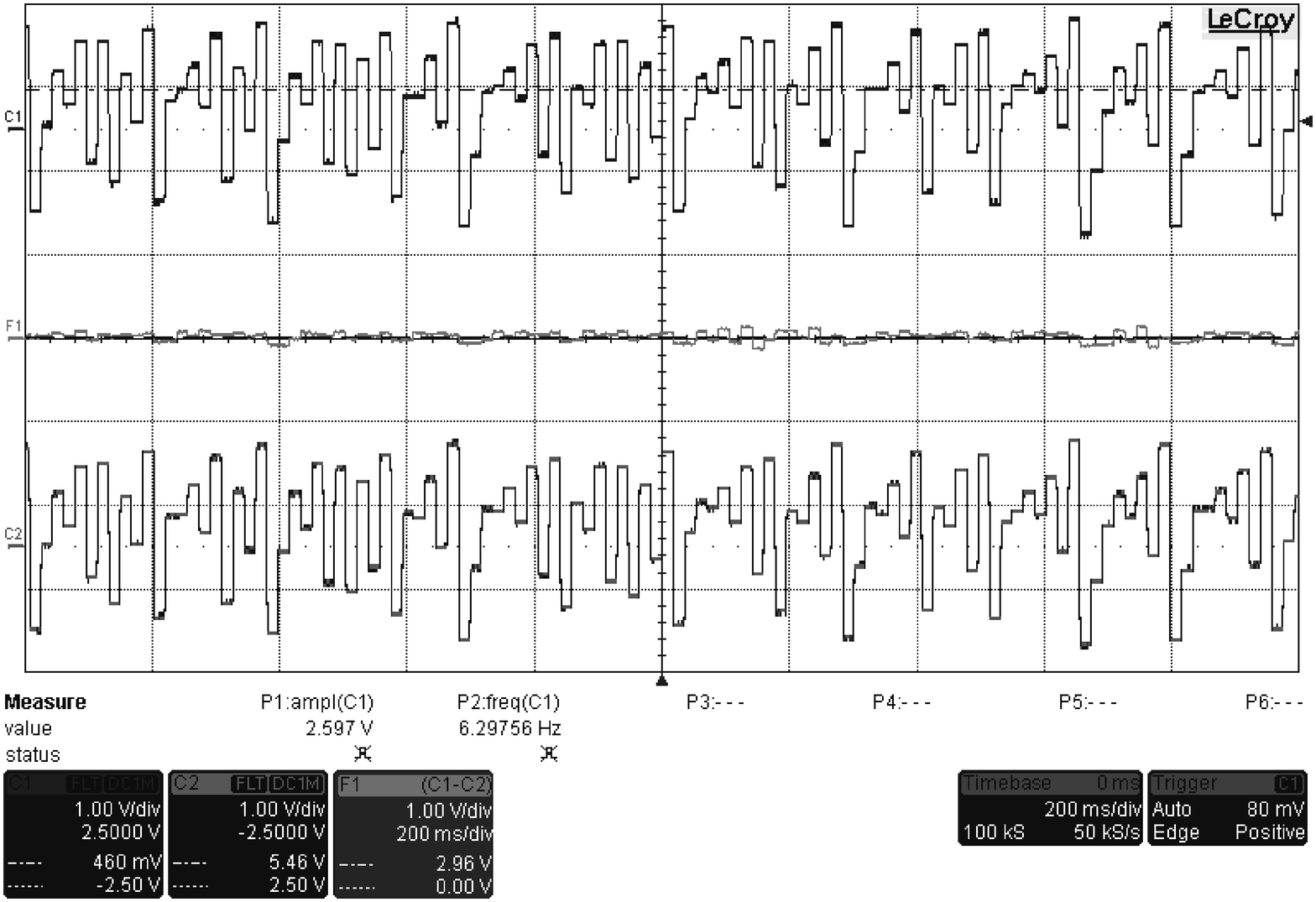}
\caption{$\alpha_c$=0.4}
\label{fig0.4}
\end{figure}

\begin{figure}
\includegraphics[scale=.25]{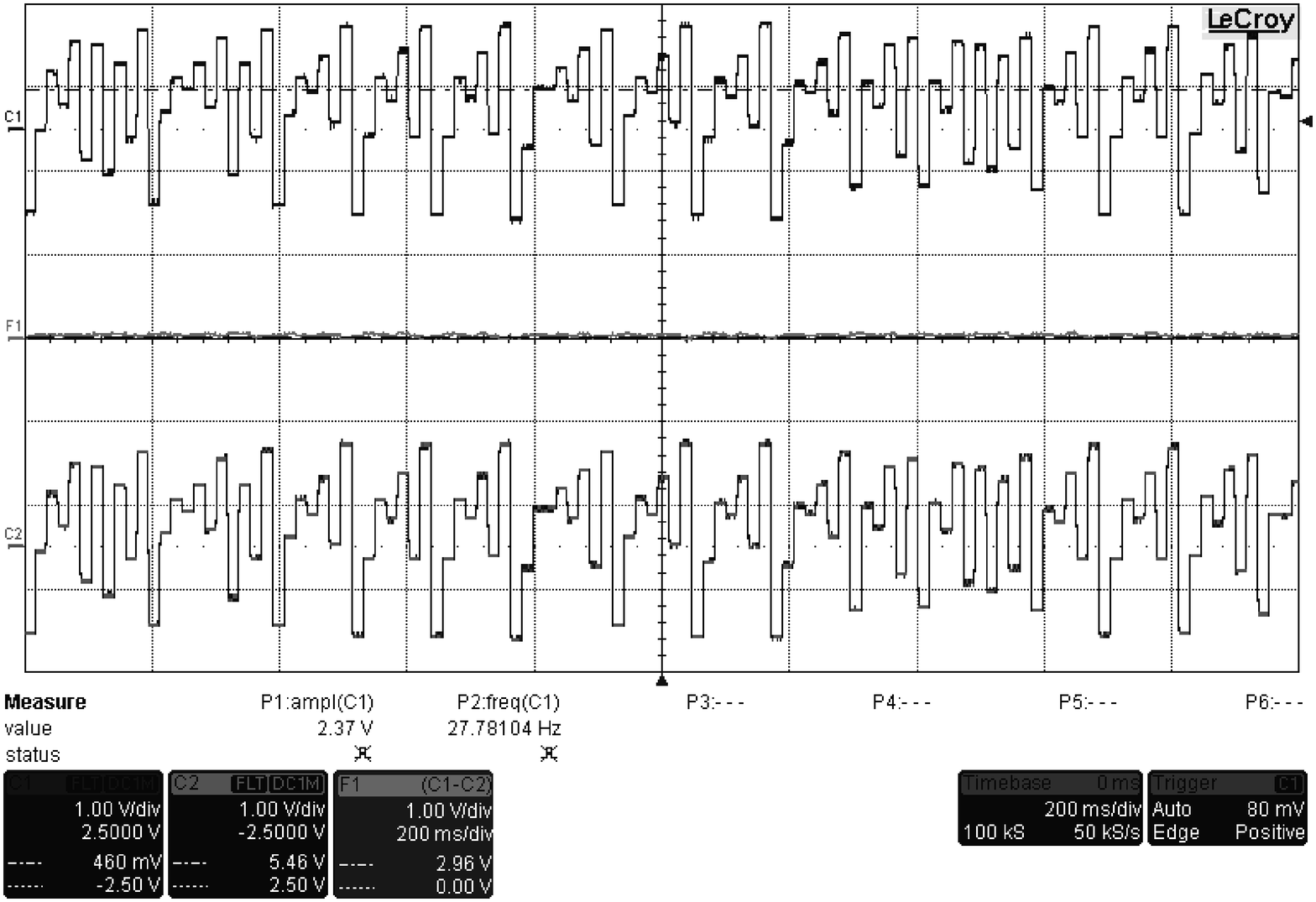}
\caption{$\alpha_c$=0.5}
\label{fig0.5}
\end{figure}

The driver and driven voltages pertaining to $x_k^r$ and $x_k^n$ were captured on a digital storage oscilloscope and diplayed as trace C1 and C2, respectively..  The error voltage pertaining to these, was also computed and displayed on the oscilloscope as trace F1.  Such oscilloscope trace, pertaining to an un-coupled pair of maps ($\alpha_c = 0$) is shown in figure \ref{UncoupledTrace}.  The lack of synchronization is evident.  Oscilloscope trace for $\alpha_c = 0.4$ is shown in figure \ref{fig0.4}.  It can be seen that the maps synchronize with a small error voltage.  Oscilloscope trace for $\alpha_c = 0.5$ is shown in figure \ref{fig0.5}.  It can be seen that the maps synchronize with an error voltage of almost zero.
\section{Potential applications}

One potential application of such electronic circuit realization of synchronized robustly chaotic maps is in cryptography. Another potential application is in secure wireless communication.

\end{document}